# QVecOpt: An Efficient Storage and Computing Optimization Framework for Large-scale Quantum State Simulation

Mingyang Yu, Haorui Yang, Donglin Wang, Desheng Kong, Ji Du, Yulong Fu and Jing Xu, *Member, IEEE*

*Abstract*—In response to the challenges in large-scale quantum state simulation on classical computing platforms, including memory limits, frequent disk I/O, and high computational complexity, this study builds upon a previously proposed hierarchical storage-based quantum simulation system and introduces an optimization framework, the Quantum Vector Optimization Framework (QVecOpt). QVecOpt integrates four strategies: amplitude pairing, cache optimization, block storage optimization, and parallel optimization. These collectively enhance state vector storage and computational scheduling. The amplitude pairing mechanism locates relevant amplitude pairs via bitwise XOR, reducing traversal complexity of single-qubit gates from $O(2^n)$ to $O(1)$. Cache optimization pre-allocates buffers and loads only required data, cutting disk I/O. Block storage optimization partitions the state vector for on-demand loading and local updates, reducing redundant access. Parallel optimization distributes the state vector across nodes for collaborative computation, achieving near-linear speedup. Complexity analysis shows that, compared with hierarchical storage simulation, the method reduces state vector traversals for single-qubit gates from $2^n$ to 1, removing the main bottleneck. It also lowers computational and I/O complexity from $O(2^n)$ to $O(2^n/C)$ and $O(2^n/B)$. In simulations of 16-29 qubits, efficiency improves nearly tenfold, breaking the memory bottleneck of existing tools and enabling high-bit quantum circuit simulations beyond traditional methods. This work provides an efficient, scalable solution for classical simulation of large-scale quantum computation with significant academic and practical value.

## I. INTRODUCTION

IN recent years, the rapid development of quantum computing has introduced a new computational paradigm for solving classically intractable high-complexity problems. Although quantum computing has entered the era of Noisy Intermediate-Scale Quantum (NISQ) devices[1], it still faces challenges such as high cost, stringent operating conditions, and limited coherence time [2]. Large-scale quantum state simulation serves as a vital tool in quantum algorithm research. It plays a critical role not only in algorithm design and verification, but also in hardware optimization, error analysis, and quantum system behavior prediction [3]. In 2019, Google's breakthrough in quantum computing made it extremely challenging to simulate quantum circuits using classical supercomputers, a phenomenon referred to as quantum supremacy [4-6]. Therefore, improving the efficiency and scalability of large-scale quantum state simulation has become a central task in the field of quantum computing.

Full-amplitude Schrödinger-state simulation [7, 8] is a general-purpose classical simulation method for quantum algorithms, circuits, and physical quantum devices. It represents quantum states using high-dimensional complex-valued vectors and updates them via quantum transformations, such as quantum gates, laser pulses, and algorithmic modules. However, traditional quantum state simulation methods based on matrix-vector multiplication face major challenges in storing and computing high-dimensional state vectors. Specifically, for a system of $n$ qubits, the state space has a dimension of $2^n$, and both the memory requirement and computational complexity grow exponentially, quickly exceeding the limits of classical computing resources [9]. In the actual simulation process, for each quantum gate operation, updating each state requires traversing the entire state vector and involving numerous complex multiplications and additions.

To address the limitations of classical computing resources, supercomputers and distributed computing platforms are commonly employed for large-scale quantum system simulation, leveraging distributed storage, high-speed interconnects, and GPU acceleration [9, 10]. For example, E. Pednault et al. proposed a tensor-slicing method combined with secondary storage to simulate random circuits with more than 49 qubits [11-13]. A. Fatima et al. introduced an optimized strategy based on data-level and thread-level parallelism along with efficient CPU cache utilization, enabling high-performance Schrödinger-state simulation of up to 64 qubits in theory [7]. Despite these advances in parallel computing and data storage, memory usage remains the primary bottleneck for simulating

This work was supported by Natural Science Foundation of Tianjin Municipality under Grant 21JCYBJC00110. (*Corresponding author: Jing Xu*.)

Mingyang Yu, Haorui Yang, Dengsheng Kong, Ji Du and Jing Xu are with the College of Artificial Intelligence, Nankai University, Tianjin, 300350, China (e-mail: 1120240312@mail.nankai.edu.cn; e-mail: 2120240645@mail.nankai.edu.cn; e-mail: kongds@mail.nankai.edu.cn; e-mail:1120230244@mail.nankai.edu.cn; e-mail: xujing@nankai.edu.cn).

Donglin Wang is with Beijing Sursen Information Technology Co., Ltd, Beijing, 100080, China (e-mail: wangdonglin@sursen.net).

Yulong Fu is with the China Academy of Electronics and Information Technology, Beijing 100041, China, and also with the Yangtze Delta Industrial Innovation Center of Quantum Science and Technology, Suzhou 215133, China (e-mail: fuyulong@cetc.com.cn).

larger quantum systems. The exponential growth in computational and storage demands significantly limits their applicability in ultra-large-scale scenarios. Therefore, it is still of great practical significance to explore more efficient storage and computation optimization strategies to improve the ability of quantum state simulation.

In our previous work [14], we developed a quantum simulation system that significantly reduced system cost by introducing a hierarchical storage architecture. However, this approach also introduced substantial challenges related to I/O overhead. The hierarchical storage architecture enables huge state vector data to be stored on cheap storage media such as hard disks. The data storage cost, which accounts for approximately 90% of the total system cost, has been reduced by an order of magnitude. Frequent data loading and refreshing lead to severe I/O bottlenecks, ultimately limiting the practicality of this approach.

To address the above challenges, this paper proposes an efficient optimization framework for large-scale Schrödinger-state simulation on hierarchical storage-based computing platforms, termed the Quantum Vector Optimization Framework (QVecOpt). This framework achieves efficient storage management, data access and computing scheduling of state vectors through the collaborative design of four strategies: single quantum gate optimization based on amplitude pairing mechanism, cache optimization, block storage optimization, and parallel optimization. The amplitude pairing mechanism exploits the locality of single-qubit gate operations to dynamically identify and couple amplitude pairs with XOR-related indices, enabling all updates to be completed in a single linear traversal and reducing computational complexity. The cache optimization strategy uses a sliding window to load only the required subset of states during computation, effectively reducing disk I/O operations. Block storage optimization partitions the state vector into structured blocks to support on-demand loading and localized updates. Parallel optimization distributes the state space across multiple computing nodes, enabling efficient resource utilization and near-linear speedup. As a result, the number of full state vector traversals is reduced from $2^n$ to 1; computational complexity decreases from $O(2^n)$ to $O(2^n/C)$, and I/O complexity is reduced from $O(2^n)$ to $O(2^n/B)$, where $C$ is the number of parallel nodes and $B$ is the number of states loaded per cache operation. Through the synergistic effect of these strategies, QVecOpt significantly reduces both computational complexity and disk I/O overhead, enabling classical platforms with limited hardware to simulate larger and more complex quantum circuits. The main contributions of this work are summarized as follows:

- A quantum state amplitude pair coupling calculation model is proposed. Through the bit-order-sensitive amplitude dynamic matching mechanism, the joint optimization of computing and storage for single-bit gate operations is achieved.
- A cache optimization strategy based on a sliding window is introduced. By pre-allocating memory buffers and loading only the required state subset, it fundamentally reduces disk access and I/O complexity.
- A block storage structure is proposed, where the large state vector is partitioned into fixed-size blocks. An index-aligned strategy reduces cross-node communication, and in combination with cache mechanisms, enables block-wise loading and computation, further enhancing locality and update efficiency.
- A parallel optimization scheme is designed to distribute the quantum state vector across multiple computing nodes. It supports collaborative execution of quantum gates, maximizes resource utilization, and achieves linear speedup under ideal conditions.

The collaborative design of the four types of optimization strategies ensures that the operation process of the quantum gate only requires one traversal of the state space to complete all computing tasks, completely solving the biggest bottleneck of the hierarchical storage architecture and providing a solid theoretical basis and engineering experience for the subsequent design and implementation of the quantum simulator.

The structure of this paper is organized as follows: Section II provides a detailed overview of traditional quantum state simulation methods and their performance bottlenecks, analyzing the primary causes of computational and I/O complexity. Section III introduces four optimization strategies to address these bottlenecks: the amplitude pairing mechanism, cache optimization, block storage optimization, and parallel optimization. The design principles and implementation details of each strategy are described. Section IV presents the experimental results, comparing the original method with the optimized version to validate the effectiveness of the proposed strategies. Finally, Section V concludes the paper and discusses future directions for improving large-scale quantum simulation systems.

## II. THE COMPUTATIONAL AND STORAGE BOTTLENECKS OF QUANTUM STATE SIMULATION

Traditional quantum computing methods typically use matrix-vector multiplication to manipulate state vector data. For a quantum system with $n$ qubits, the state space has a dimension of $2^n$, and the corresponding state vector requires substantial memory. When simulating quantum computation on classical hardware, the quantum state of an $n$-qubit system is typically represented using a dense vector. The state vector $|\psi\rangle$ expressed as:

$$|\psi\rangle = \begin{bmatrix} \alpha_0 \\ \alpha_1 \\ \vdots \\ \alpha_{2^n-1} \end{bmatrix}, \quad \alpha_i \in \mathbb{C} \qquad (1)$$

where, each amplitude $\alpha_i$ is a complex number. In double-precision format, it requires 16 bytes of storage, while in single-precision format, it takes 8 bytes. Therefore, the total storage size of the full state vector is:

$$S_{\text{total}} = 2^n \times S_{\text{state}}, \qquad (2)$$

where, $S_{\text{state}}$ denotes the storage size of each $\alpha_i$, which is 16 bytes in double precision or 8 bytes in single precision.

The exponential growth of the state vector leads to extremely high storage costs in large-scale quantum state simulation, limiting its practical applicability. To address this issue, our previous work [14] proposed a hierarchical storage architecture for quantum simulation. This architecture prioritizes the deployment of computing nodes in high-performance storage areas such as memory to ensure computing performance. The complete state vector is stored on low-cost storage media such as disks, significantly reducing the overall system cost.

However, this hierarchical storage architecture also introduces a new performance bottleneck—frequent I/O operations on low-cost storage devices such as disks. Specifically, simulating a single quantum gate requires multiple traversals of the state vector, with traversal counts growing exponentially. This leads to significant I/O overhead, which becomes the primary limitation on overall simulation performance. To address this issue, this paper further proposes a set of optimization strategies based on the hierarchical storage architecture, aiming to reduce access frequency to low-cost storage and improve data scheduling efficiency, thereby enhancing the overall execution performance of quantum simulation tasks under this architecture.

In traditional quantum state simulation, although basic quantum gates (such as Hadamard and CNOT) intrinsically operate on only one or two qubits. However, in the simulation process, it is still usually extended to a complete $2^n \times 2^n$ dimensional complex matrix U by Kronecker (tensor) product with $n-1$ identity $2 \times 2$ identity matrices. This matrix is then used to update the state of the entire $n$-qubit system. The gate matrix $U$ role in the current system state vector $|\psi\rangle$, update process expression is as follows:

$$|\psi'\rangle = U|\psi\rangle, \quad (3)$$

$|\psi'\rangle$ is the new state vector, where $i$th component is:

$$\alpha'_i = \sum_{j=0}^{2^n-1} U_{ij}\, \alpha_j. \quad (4)$$

It is evident that computing each updated amplitude $\alpha'_i$ requires accessing and summing over $2^n$ complex numbers, resulting in $2^n \times 2^n$ complex multiplications and additions. Therefore, the computational complexity of a single quantum gate operation reaches $O(2^n \times 2^n) = O(2^{2n})$. Since updating each $\alpha'_i$ requires a full traversal of the quantum state vector, each gate operation involves $2^n$ traversals. As a result, the state vector becomes the primary bottleneck limiting the scalability of large-scale simulation.

In summary, the hierarchical storage architecture exhibits the following performance bottlenecks:

- Exponential state vector traversal: Each quantum gate operation requires up to $2^n$ traversals of the state vector, resulting in a traversal complexity of $O(2^n)$. Under the hierarchical storage architecture, where the state vector resides on low-performance storage devices such as hard disks, the cost of such frequent access becomes prohibitive.
- High I/O throughput: The data volume per quantum gate operation reaches $2^n \times 2^n \times S_{\text{state}} = 2^{2n} \times S_{\text{state}}$, resulting in an I/O complexity of $O(2^{2n})$. This immense I/O throughput imposes excessive pressure on the storage system.

As quantum computing scales up, the storage and computational complexity of high-dimensional state vectors has emerged as a primary bottleneck. In classical methods, each gate operation requires $2^n$ full traversals of the state vector of length $2^n$, resulting in exponential growth in both I/O and computational complexity. Although the previously proposed hierarchical storage approach addresses the limitations of memory capacity and cost, it introduces disk I/O bottlenecks that severely restrict overall performance.

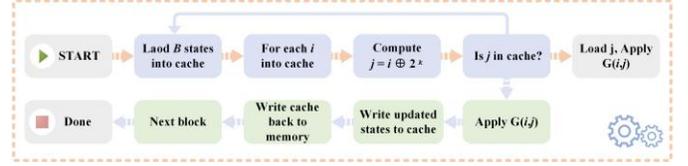

**Fig. 1.** Cache Scheduling flowchart.

### III. DESIGN AND PRINCIPLE OF OPTIMIZATION METHODS

To address the above issues, this work proposes four strategies based on the hierarchical storage architecture: amplitude pairing mechanism, cache optimization, block storage optimization, and parallel optimization. These strategies jointly reduce both computational and I/O complexity. The number of state vector traversals per gate operation is reduced from $2^n$ to 1, significantly decreasing disk I/O overhead. As a result, the proposed approach substantially improves the computational efficiency and scalability of large-scale quantum state simulation.

#### A. Single Quantum Gate Optimization Based on Amplitude Pairing

As discussed in Section II, traditional quantum state simulation methods typically perform single-qubit gate operations via dense matrix multiplication, resulting in a computational complexity of $O(2^{2n})$ per gate and requiring $2^n$ full traversals of the state vector, which severely limits simulation efficiency. To overcome these bottlenecks, QVecOpt introduces an optimization strategy based on the amplitude pairing mechanism, which exploits the locality of single-qubit gates to significantly reduce computational complexity and traversal overhead.

For a single-qubit gate acting on the $k$th qubit, it only affects amplitude pairs $(i, i \oplus 2^k)$ in the state vector, where $\oplus$ denotes the bitwise XOR operation. The amplitude update process can be locally expressed as:

$$\begin{pmatrix} \alpha'_i \\ \alpha'_{i\oplus 2^k} \end{pmatrix} = U \begin{pmatrix} \alpha_i \\ \alpha_{i\oplus 2^k} \end{pmatrix}, \quad (5)$$

where, $U$ presents a $2 \times 2$ single-qubit gate matrix, and $\alpha_i$ and $\alpha_{i\oplus 2^k}$ denote the amplitudes at the corresponding indices in the original state vector. Specifically, during traversal, the simulator sequentially scans index $i$ of the state vector and checks the bit value at position of index $i$. If the bit value at position of index $i$ is 0, index $i$ is paired with $i \oplus 2^k$, and the simulator applies the local single-qubit gate $U$ to update the amplitude pair $(\alpha_i, \alpha_{i\oplus 2^k})$ via a linear transformation. This is because if the bit value at index $i$ is 0, then $i \oplus 2^k$ must have a 1 at that bit, ensuring that each amplitude pair is processed only once, thereby avoiding redundant computation.

With the amplitude pairing update mechanism, each amplitude pair is accessed only once, eliminating redundant reads and computations. The total number of pairs is $2^{n-1}$, and the corresponding single quantum gate operation only needs to sequentially traverse the entire state vector once, which not only reduces the total computational complexity from $O(2^{2n})$ to $O(2^n)$, but also reduces the traversal complexity from $O(2^n)$ to $O(1)$.

In summary, the amplitude pairing update mechanism restructures the execution path of traditional simulation by enabling unified gate operations within a single traversal of the state vector. This reduces the traversal complexity from $O(2^n)$ to a single pass, representing a fundamental theoretical advancement in the framework. To ensure efficient execution of this mechanism in large-scale systems, QVecOpt incorporates sliding-window state caching, block storage partitioning, and parallel distribution strategies. These enhancements provide robust support from both storage management and computational resource scheduling perspectives.

*B. Sliding-Window State Caching Mechanism*

As described in Section III.A, the amplitude pairing update mechanism enables all amplitude updates for a single-qubit gate to be completed through a single sequential traversal of the full state vector. Building upon this mechanism, QVecOpt introduces a Sliding-Window State Caching strategy to further alleviate I/O overhead and improve access efficiency. The proposed sliding-window mechanism pre-allocates a cache region of size $M$ in memory and loads only the state vector blocks relevant to the current computation, thereby reducing the frequency of disk accesses. The number of states that can be cached in a single pass is defined as:

$$B = \frac{M}{S_{\text{state}}}, \tag{6}$$

where, $S_{\text{state}} = 16$ bytes indicates that each quantum amplitude is stored using the complex128 format, occupying 16 bytes; $B$ denotes the maximum number of amplitudes that can be loaded into memory at once during computation. With the cache optimization strategy, although the overall computational complexity remains $O(2^{2n})$, the I/O complexity is significantly reduced to $O(2^n/B)$ by avoiding full state vector loading in each computation cycle. During each gate operation, only a localized subset of the full state space is accessed, avoiding exhaustive $2^n$ disk loading and thereby saving substantial I/O cost.

The core of this optimization lies in transforming gate computation into an "on-demand loading + in-cache computation + local write-back" process, breaking away from the traditional "full-load full-compute" model and greatly improving the simulator's ability to handle high-qubit quantum states. In addition, the sliding-window state caching mechanism lays the foundation for subsequent block storage and parallel optimization strategies. The cache scheduling process is illustrated in Fig. 1.

*C. Block Storage Optimization*

Although the sliding-window mechanism effectively reduces disk I/O during a single sequential traversal, as the number of qubits increases, the exponential growth of the state vector causes the size of each cache load to become a limiting factor for system throughput. To further improve data locality and cache utilization, QVecOpt introduces a block-based storage optimization strategy on top of the sliding-window caching mechanism. The state vector is partitioned into fixed-size blocks, and during computation, only the blocks affected by the current gate are loaded, thereby further reducing read/write operations. $S_{\text{block}}$ denote the size of each block in bytes. Then, the state vector is divided into:

$$N_{\text{blocks}} = \frac{S_{\text{total}}}{S_{\text{block}}} = \frac{2^n \times S_{\text{state}}}{S_{\text{block}}}, \tag{7}$$

where, $N_{\text{blocks}}$ denotes the total number of blocks into which the quantum state vector is partitioned. Each block contains $B$ consecutive quantum amplitudes.

In standard algorithms, each quantum gate operation traverses and updates the entire $2^n$-dimensional state vector. However, in practice, many quantum gates (e.g., single-qubit and two-qubit gates) only affect transformations between pairs of amplitudes, meaning that only a portion of the state space needs to be accessed per operation. By adopting a block-partitioned storage structure, the simulator loads only the blocks affected by a given gate operation, avoiding full-state traversal. As a result, the I/O complexity is reduced from the original $O(2^n)$ to $O(2^n/B)$.

In summary, block storage optimization provides structural support for sliding-window state cache optimization. In the in-memory cache, blocks are loaded one by one instead of individual states, thereby achieving "block-level cache management". Assuming the total cache size is $M$, the number of blocks that can be loaded in a single pass is:

$$N_{\text{cache\_blocks}} = \frac{M}{S_{\text{block}}}. \tag{8}$$

The caching mechanism is responsible for managing the loading, computation, and write-back of these blocks in memory, while the block storage mechanism ensures that these blocks are well-organized and efficiently accessible on disk.

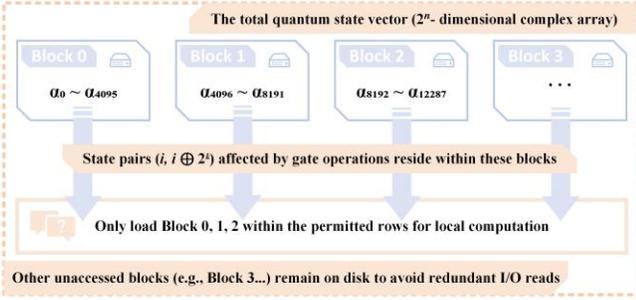

Fig. 2. Block-based quantum state storage and selective data access.

Block storage optimization partitions the quantum state vector into multiple data blocks, enabling on-demand loading and block-wise access, which significantly reduces reading overhead of irrelevant data. This approach supports localized updates of the state vector, avoids redundant I/O operations, and works in tandem with the caching mechanism to improve memory scheduling efficiency. Moreover, data blocks serve as natural computational units, facilitating efficient task partitioning for parallel processing. Overall, block storage optimization effectively mitigates the storage bottlenecks in high-qubit state simulation and constitutes a key component in building high-performance quantum simulation systems. The block storage structure and controlled data access process are illustrated in Fig. 2, where the state vector is partitioned into fixed-size data blocks. Each block contains a sequence of complex-valued amplitudes (e.g., Block 0 is loaded into the cache and participates in computation, while other blocks remain unread). This approach significantly reduces disk I/O operations.

### D. Parallel Computation Optimization

Although sliding-window and block storage optimizations significantly improve large-scale state vector processing within a single node, the rapid growth in qubit count eventually overwhelms the computational and memory resources of a single machine, making it insufficient for simulating ultra-large quantum circuits. To address this, QVecOpt introduces a parallel computation optimization strategy based on single-pass traversal and localized data partitioning. The quantum state vector is divided and distributed across multiple computing nodes, which collaboratively execute quantum gate operations.

Assume the system consists of $C$ computing nodes. The entire $2^n$-dimensional quantum state vector is evenly partitioned into $C$ subranges, with each node responsible for a contiguous range of state indices. The task interval assigned to the $i$th node is defined as:

$$\left[\frac{2^n \times i}{C}, \frac{2^n \times (i+1)}{C} - 1\right], 0 \leq i < C. \quad (9)$$

Each node independently loads its local state blocks and performs quantum gate computations within its assigned range. Ideally, both the total computational and I/O complexity can be reduced linearly with respect to the number of nodes, reaching $O(2^n/C)$. The structure of state vector partitioning is illustrated in Fig. 3. In a parallel computing environment, the state vector is evenly divided into $C$ segments, each assigned to a different computing node. Each node is responsible for gate computations within its local segment, theoretically reducing both computational and I/O complexity to $O(2^n/C)$.

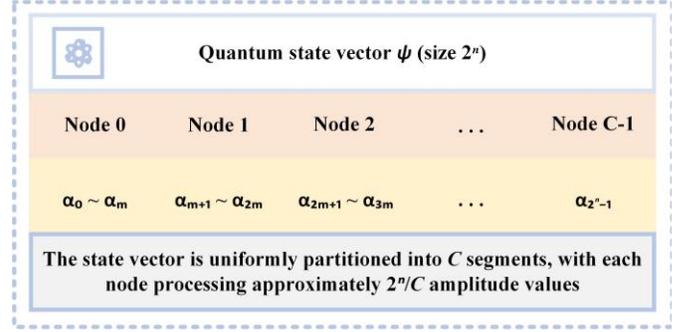

Fig. 3. State vector partitioning across $C$ computing nodes.

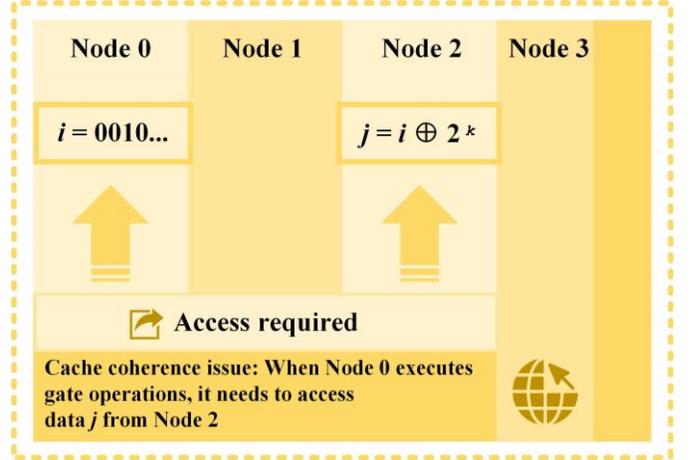

Fig. 4. Cross-node amplitude pair dependencies.

Although the state vector can theoretically be evenly partitioned across nodes, single-qubit gate operations involve amplitude pairs $(i, i \oplus 2^k)$. If such pairs span across node boundaries, they may introduce cache fragmentation and data dependencies, resulting in load imbalance. Specifically, when the cache size $M$ is insufficient, the paired indices $(i, i \oplus 2^k)$ may reside on different nodes. In this case, some nodes need to access non-local state blocks or undertake repeated calculations of amplitude pairs across nodes, resulting in local load doubling. Under such conditions, the overall computational complexity increases to $O(2^{n+1}/C)$. Cross-node dependencies between amplitude pairs $(i, i \oplus 2^k)$ are illustrated in Fig. 4. Some state index pairs $(i, i \oplus 2^k)$ may belong to different nodes, resulting in data dependence between nodes and need to communicate across nodes. Without index-aligned scheduling or data prefetching, such dependencies may cause cache fragmentation, degrading parallel efficiency and load balance. Therefore, in practical implementations, careful task partitioning is required to minimize cross-node interactions and maintain computational balance.

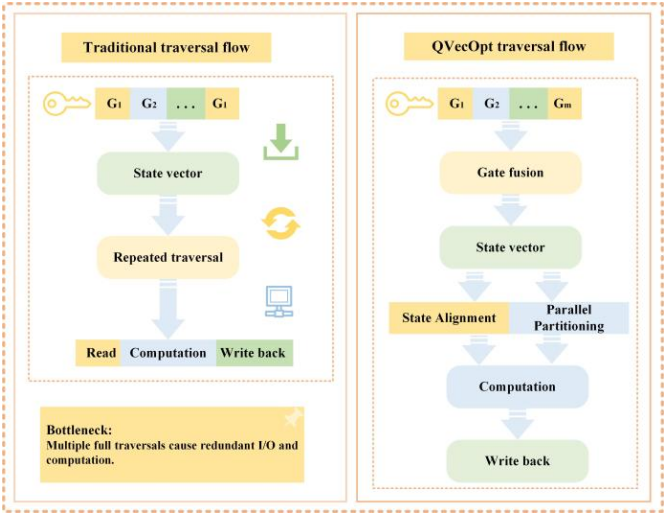

**Fig. 5.** Comparison between the traditional traversal process and the QVecOpt single traversal process.

TABLE I
COMPLEXITY COMPARISON OF OPTIMIZATION STRATEGIES

| Optimizing strategy | Computation complexity | I/O complexity [a] | Traversal complexity |
|---|---|---|---|
| Original method | $O(2^{2n})$ | $O(2^{2n})$ | $O(2^n)$ |
| Amplitude pairing mechanism | $O(2^n)$ | $O(2^n)$ | $O(1)$ |
| Cache optimization | $O(2^n)$ | $O(2^n/B)$ | $O(1)$ |
| Block storage optimization | $O(2^n)$ | $O(2^n/B)$ | $O(1)$ |
| Parallel optimization (ideal) | $O(2^n/C)$ | $O(2^n/C)$ | $O(1)$ |
| Parallel optimization (unbalanced) | $(2^{n+1}/C)$ | $(2^{n+1}/C)$ | $O(1)$ |

In Table I, $B$ denotes the number of quantum states that can be held in cache, and $C$ represents the number of parallel computing nodes.

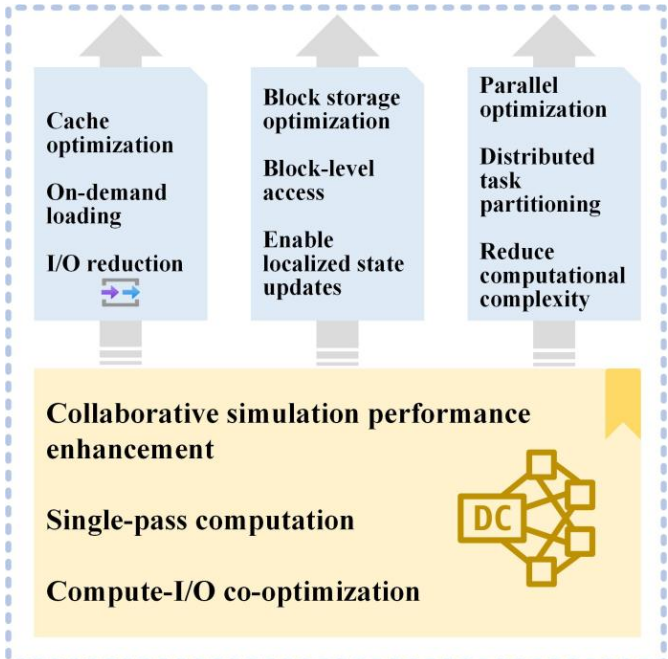

**Fig. 6.** Diagram of the collaborative relationship of optimization strategies.

*E. Analysis and Discussion*

Based on the optimizations presented in Sections III.A through III.D, QVecOpt achieves a fundamental breakthrough over traditional quantum state simulation workflows. The core improvement lies in the amplitude pairing update mechanism, which optimizes the traversal mode of the state vector from the traditional "each state vector operation needs to be traversed once" to "the entire quantum gate needs to be traversed once uniformly".

Specifically, traditional methods require a full traversal of all $2^n$ amplitudes for each gate operation, resulting in a traversal complexity of $O(2^n)$. In QVecOpt, all gate operations are batch-processed during a single sequential scan of the state vector, reducing the number of traversals to $O(1)$. While maintaining the ability to process $2^n$ amplitudes per traversal, this approach significantly improves the overall efficiency and scalability of the simulation workflow. As illustrated in Fig. 5, QVecOpt enables a paradigm shift from traditional multi-pass simulation to a single-pass model through a unified traversal strategy.

Building on the traversal reduction mechanism, QVecOpt further integrates three system-level optimization strategies: sliding-window state caching, block storage optimization, and parallel computation. These strategies collectively enhance system performance from the perspectives of memory I/O management, storage layout, and computational resource distribution. As shown in Fig. 6.

Firstly, the amplitude pairing mechanism uses the local characteristics of quantum gate operation to dynamically couple the amplitude pairs satisfying the XOR relationship in a single sequential traversal. This reduces traversal complexity from $O(2^n)$ to $O(1)$ and computational complexity from $O(2^{2n})$ to $O(2^n)$, compared to traditional dense matrix-vector multiplication. Moreover, it transforms the memory access pattern from random to sequential, laying the foundation for further performance enhancements.

Secondly, the sliding-window state caching mechanism uses a Sliding Window to dynamically load local data blocks in a single sequential traversal. This significantly reduces the frequency of disk I/O operations. With appropriate cache size $B$, the I/O complexity is reduced from $O(2^n)$ to $O(2^n/B)$.

Block storage optimization partitions the quantum state vector into structured blocks and aligns the amplitude pairs $(i, i \oplus 2^k)$ locally in the partition process to maximize local access continuity and cache utilization. This block-wise partitioning further reduces I/O complexity to $O(2^n/M)$, where $M$ denotes the number of amplitudes contained in a single data block.

Finally, Parallel Computation Optimization builds upon the block partitioning strategy by evenly distributing the state vector across $C$ computing nodes. Each node performs sequential local traversal and independently updates its assigned amplitude pairs, reducing the overall computational complexity to $O(2^n/C)$. Leveraging the unified traversal strategy, all nodes can proceed in parallel without incurring global synchronization overhead, thereby enhancing both scalability and throughput of the system.

Table I summarizes the computational complexity, I/O complexity, and number of state vector traversals under different optimization strategies. The results demonstrate that with the introduction of each optimization strategy, both computational and storage complexity are significantly reduced, leading to a substantial improvement in overall system throughput.

TABLE II
RUN TIME COMPARISON OF DIFFERENT METHODS

| Qubits | Data Size | Qiskit | QVecOpt (Single-threaded) | QVecOpt (Dual-threading) |
|---|---|---|---|---|
| 16 | 1M | 0.19 s | 0.05 s | 0.05 s |
| 20 | 16M | 0.97 s | 0.11 s | 0.09 s |
| 24 | 256M | 13.31 s | 0.82 s | 0.65 s |
| 25 | 512M | 37.00 s | 1.55 s | 1.16 s |
| 26 | 1G | 67.74 s | 3.00 s | 2.37 s |
| 27 | 2G | 143.75 s | 6.17 s | 4.74 s |
| 28 | 4G | Failed (OOM) | 9.99 s | 7.13 s |
| 29 | 8G | Failed (OOM) | 19.36 s | 14.55 s |

As shown in Table I, with the incremental introduction of optimization strategies, QVecOpt maintains the single-pass traversal model while reducing I/O complexity from the original $O(2^n)$ to $O(2^n/B)$ through the combined effect of cache and block storage optimizations. With the addition of parallel execution, the computational complexity is further reduced to $O(2^n/C)$, significantly decreasing both computation and I/O overhead while enabling strong distributed scalability. In particular, the unified traversal strategy in QVecOpt overcomes the fundamental limitation of repeated state vector traversals per gate in traditional methods, providing a robust architectural foundation for simulating ultra-large-scale quantum circuits.

The above analysis and calculations are based on full-amplitude quantum simulation, but the proposed optimization strategies are also applicable to partial-amplitude simulation and tensor network-based quantum simulation.

## IV. EXPERIMENTAL EVALUATION

### A. Experimental Setup

All experiments in this paper were conducted on a computing platform equipped with an Intel Core i7-14900HX processor and 64 GB of memory, running Windows 11 (version 23H2). The Qiskit framework used for performance comparison was version 1.3.1, and it was executed natively in the Windows environment. The proposed optimization code was deployed within the Windows Subsystem for Linux environment on the same Windows 11 system. Within the WSL environment, an 8 GB in-memory disk was mounted as temporary storage to further reduce the impact of disk I/O on experimental results. All test data and code were executed uniformly in the described environment to ensure fairness and consistency in the experiments. The experiments were performed on quantum systems with 16 to 29 qubits, corresponding to data sizes ranging from 1 MB to 8 GB.

### B. Performance Metrics and Evaluation Methodology

To comprehensively and objectively evaluate the proposed optimization strategies, we adopt two primary performance metrics: Execution Time and Scalability. Specifically, execution time is defined as the total elapsed time from the start of quantum gate execution to the completion of state vector updates, which reflects the degree of performance improvement achieved by the optimization strategies during actual computation. Scalability characterizes the growth trend of execution time as the number of qubits increases, which indicates the effectiveness and potential limitations of each optimization strategy in large-scale quantum simulation scenarios.

In the experimental evaluation, we selected Qiskit, a widely adopted quantum simulation framework, as the performance baseline. It is comprehensively compared against the proposed optimization method, evaluated in both single-threaded and multi-threaded modes. All experiments were conducted under identical hardware and software environments to ensure fairness and accuracy of the evaluation. By quantitatively comparing the above metrics and analyzing their trends, the experimental results objectively demonstrate the effectiveness of the proposed method in terms of computational efficiency and scalability.

### C. Experimental Results and Analysis

The experimental results are shown in Table II, which lists the running time of each method with different number of qubits.

As shown in Table II, the proposed optimization method significantly outperforms Qiskit in terms of execution efficiency. The performance advantage becomes increasingly pronounced at medium to high qubit counts (24 to 29 qubits). When the number of qubits reaches 28 or more, Qiskit fails to operate normally due to memory bottlenecks. However, the method proposed in this paper still maintains stable and efficient operation, fully verifying the effectiveness of caching, block storage and parallel computing strategies.

### D. Performance Comparison and Discussion

A further comparison between the single-threaded and multi-threaded modes reveals additional performance gains under multi-threading. For example, in the case of 29 bits, the running time of the single-threaded mode is 19.36 seconds, and the running time of the two-threaded mode is reduced to 14.55 seconds, a performance improvement of about 25%. This trend is expected to become even more prominent for larger-scale simulation tasks.

Moreover, the proposed method demonstrates particularly notable improvements in storage and I/O efficiency. Unlike Qiskit, which suffers from memory exhaustion due to loading the entire state vector at once, the proposed cache and block storage optimizations enable simulation of much larger quantum systems. The parallel optimization strategy further enhances the system's overall computational capacity and scalability, exhibiting a strong and consistent performance scaling trend.

In summary, the experimental results demonstrate that the proposed method significantly enhances the performance of classical quantum state simulation and effectively addresses the memory and I/O bottlenecks encountered in large-scale scenarios, making it well-suited for simulating medium to large-scale quantum circuits.

## V. Conclusion

Building upon a previously proposed hierarchical storage architecture [14], this work further introduces a systematic set of optimization framework to address the challenges of high memory pressure, frequent I/O operations, and computational complexity when simulating large-scale quantum circuits, thereby enhancing overall simulation performance. The framework first uses a paired amplitude pairing update mechanism to realize that a single quantum gate operation completes all calculations in a single sequential traversal of the state vector, and optimizes the high-frequency traversal mode in which each gate operation in the traditional method needs to fully traverse the $2^n$ amplitude to a unified single traversal mode for the entire circuit.

We further designed three additional methods, including the sliding-window state caching mechanism, block storage optimization, and parallel computation optimization, in order to achieve efficient storage management, data access, and computational scheduling for the quantum state vector. The sliding-window state caching mechanism reduces disk I/O operations by dynamically loading only the relevant portions of the state vector during sequential traversal. Block storage optimization partitions the state vector into structured data blocks, which significantly reduces redundant data access and improves locality. Parallel computing optimization significantly improves the overall simulation performance and system scalability by reasonably dividing the state space and executing it in parallel on multiple computing nodes. Without modifying the algorithmic logic or simulation accuracy, the optimized method reduces the number of state vector traversals from $2^n$ to just one. The computational complexity is reduced from $O(2^n)$ to $O(2^n/C)$, and the I/O complexity is reduced from $O(2^n)$ to $O(2^n/B)$, where $C$ represents the number of parallel computing nodes and $B$ denotes the number of quantum states loaded per cache window.

Experimental results show that the optimization strategy proposed in this paper has obvious performance advantages over existing quantum simulation tools (such as Qiskit). In simulation tests with a scale of 16 to 29 qubits, our method not only significantly reduces the running time (up to nearly 10 times), but also breaks through the memory limitations of Qiskit in larger-scale scenarios. It is capable of successfully simulating high-qubit quantum circuits that Qiskit fails to execute. In addition, further multithreaded optimization yields an additional performance gain of approximately 25%, demonstrating good scalability.

In summary, the optimization scheme proposed in this paper significantly improves the performance of classical quantum state simulation in terms of computing and storage, effectively solves the key bottleneck problems faced by traditional methods. This work provides a theoretical basis and engineering practice experience for building an efficient and scalable classical quantum simulation platform, and has important academic value and application prospects.

## References

[1] J. Preskill, "Quantum computing in the NISQ era and beyond," *Quantum*, vol. 2, pp. 79, 2018.

[2] K. C. Miao, J. P. Blanton, C. P. Anderson, A. Bourassa, A. L. Crook, G. Wolfowicz, H. Abe, T. Ohshima, and D. D. Awschalom, "Universal coherence protection in a solid-state spin qubit," *Science*, vol. 369, no. 6510, pp. 1493-1497, 2020.

[3] K. De Raedt, K. Michielsen, H. De Raedt, B. Trieu, G. Arnold, M. Richter, T. Lippert, H. Watanabe, and N. Ito, "Massively parallel quantum computer simulator," *Computer Physics Communications*, vol. 176, no. 2, pp. 121-136, 2007.

[4] S. Boixo, S. V. Isakov, V. N. Smelyanskiy, R. Babbush, N. Ding, Z. Jiang, M. J. Bremner, J. M. Martinis, and H. Neven, "Characterizing quantum supremacy in near-term devices," *Nature Physics*, vol. 14, no. 6, pp. 595-600, 2018.

[5] F. Arute, K. Arya, R. Babbush, D. Bacon, J. C. Bardin, R. Barends, R. Biswas, S. Boixo, F. G. Brandao, and D. A. Buell, "Quantum supremacy using a programmable superconducting processor," *Nature*, vol. 574, no. 7779, pp. 505-510, 2019.

[6] A. P. Lund, M. J. Bremner, and T. C. Ralph, "Quantum sampling problems, BosonSampling and quantum supremacy," *npj Quantum Information*, vol. 3, no. 1, pp. 15, 2017.

[7] A. Fatima, and I. L. Markov, "Faster schrödinger-style simulation of quantum circuits." pp. 194-207.

[8] L. Burgholzer, H. Bauer, and R. Wille, "Hybrid Schrödinger-Feynman simulation of quantum circuits with decision diagrams." pp. 199-206.

[9] S. Shivam, C. W. von Keyserlingk, and S. L. Sondhi, "On classical and hybrid shadows of quantum states," *SciPost Physics*, vol. 14, no. 5, pp. 094, 2023.

[10] H. De Raedt, F. P. Jin, D. Willsch, M. Willsch, N. Yoshioka, N. Ito, S. J. Yuan, and K. Michielsen, "Massively parallel quantum computer simulator, eleven years later," *Computer Physics Communications*, vol. 237, pp. 47-61, Apr, 2019.

[11] E. Pednault, J. A. Gunnels, G. Nannicini, L. Horesh, and R. Wisnieff, "Leveraging secondary storage to simulate deep 54-qubit sycamore circuits," *arXiv preprint arXiv:1910.09534*, 2019.

[12] E. Pednault, J. A. Gunnels, G. Nannicini, L. Haoresh, T. Magerlein, E. Solomonik, E. W. Draeger, E. T. Holland, and R. Wisnieff, *Breaking the 49-qubit barrier in the simulation of quantum circuits*, Lawrence Livermore National Laboratory (LLNL), Livermore, CA (United States), 2018.

[13] A. Tabuchi, S. Imamura, M. Yamazaki, T. Honda, A. Kasagi, H. Nakao, N. Fukumoto, and K. Nakashima, "mpiQulacs: A Scalable Distributed Quantum Computer Simulator for ARM-based Clusters." pp. 959-969.

[14] B. Wang, Z. Zhang, P. Siarry, X. Liu, G. Królczyk, D. Hua, F. Brumercik, and Z. Li, "A nonlinear African vulture optimization algorithm combining Henon chaotic mapping theory and reverse learning competition strategy," *Expert Systems with Applications*, vol. 236, pp. 121413, 2024.